\documentclass[useAMS,usenatbib,usegraphicx]{mn2e}
\usepackage{times}
\usepackage{amsmath}
\usepackage{amssymb}

\newcommand{\rsun}{\,\mbox{$\rm R_{\odot}$}}

\newcommand{\msun}{\,\mbox{$\rm M_{\odot}$}}
\newcommand{\mearth}{\,\mbox{$\rm M_{\oplus}$}}

\newcommand{\kms}{\hbox{kms$^{-1}$}}
\newcommand{\ms}{\hbox{ms$^{-1}$}}
\newcommand{\cms}{\hbox{cms$^{-1}$}}
\newcommand{\vsini}{\hbox{$v$\,sin\,$i$}}

\newcommand{\degs}{$\degr$}
\newcommand{\chisq}{$\chi^{2}$}

\newcommand{\ha}{H$\alpha$}

\title[Low mass M dwarf planets with ROPS]{ROPS: A New Search for Habitable Earths in the Southern Sky}
\author[J.R.~Barnes et al.]
{J.R.~Barnes$^{1}$, 
J.S.~Jenkins$^{2}$,  
H.R.A.~Jones$^{1}$,  
P. Rojo$^{2}$,  
P. Arriagada$^{3}$, 
A. Jord\'an$^{3}$,  \newauthor 
D. Minniti$^{3,4}$,  
M. Tuomi$^{1,5}$, 
S.V. Jeffers$^{6}$,
and D. Pinfield$^{1}$ \\
$^{1}$ Centre for Astrophysics Research,. University of Hertfordshire,. College Lane, Hatfield. Herts. AL10 9AB. UK. \\
$^{2}$ Departamento de Astronom\'{i}a, Universidad de Chile, Camino del Observatorio 1515, Las Condes, Santiago. Chile. \\
$^{3}$ Dept. of Astronomy and Astrophysics, Pontificia Universidad Catolica de Chile, Av. Vicuna Mackenna 4860, Santiago, Chile \\
$^{4}$ Vatican Observatory, V00120 Vatican City State, Italy \\
$^{5}$ University of Turku, Tuorla Observatory, Department of Physics and Astronomy, V\"ais\"al\"antie 20, FI-21500, Piikki\"o, Finland \\
$^{6}$ Institut f\"{u}r Astrophysik, Georg-August-Universit\"{a}t, Friedrich-Hund-Platz 1, Friedrich-Hund-Platz 1, D-37077 Göttingen. Germany.}

\begin{document}

\date{Submitted March 2012; Accepted May 2012}

\pagerange{\pageref{firstpage}--\pageref{lastpage}} \pubyear{2010}

\maketitle

\protect\label{firstpage}

\begin{abstract}
We present the first results from our Red Optical Planet Survey (ROPS) to search for low mass planets orbiting late type dwarfs (M5.5V\,-\,M9V) in their habitable zones (HZ). Our observations, with the red arm of the {\sc mike} spectrograph (0.5\,-\,\hbox{0.9 \micron}) at the 6.5 m Magellan Clay telescope at Las Campanas Observatory indicate that $\geq$ 92\% of the flux lies beyond \hbox{0.7 \micron}. We use a novel approach that is essentially a hybrid of the simultaneous iodine and ThAr methods for determining precision radial velocities. We apply least squares deconvolution to obtain a single high S/N ratio stellar line for each spectrum and cross correlate against the simultaneously observed telluric line profile, which we derive in the same way.

Utilising the 0.62\,-\,0.90 \micron\ region, we have achieved an r.m.s. precision of \hbox{10 \ms} for an M5.5V spectral type star with spectral S/N $\sim$\,160 on 5 minute timescales. By M8V spectral type, a precision of $\sim$\,30 \ms\ at S/N = 25 is suggested, although more observations are needed. An assessment of our errors and scatter in the radial velocity points hints at the presence of stellar radial velocity variations. Of our sample of 7 stars, 2 show radial velocity signals at 6$\sigma$ and 10$\sigma$ of the cross correlation uncertainties. We find that chromospheric activity (via \ha\ variation) does not have an impact on our measurements and are unable to determine a relationship between the derived photospheric line profile morphology and radial velocity variations without further observations. If the signals are planetary in origin, our findings are consistent with estimates of Neptune mass planets that predict a frequency of 13\,-\,27 per cent for {\em early} M dwarfs.

Our current analysis indicates the we can achieve a sensitivity that is equivalent to the amplitude induced by a 6 M$_\oplus$ planet orbiting in the habitable zone. Based on simulations, we estimate that $<$10 M$_\oplus$ habitable zone planets will be detected in a new stellar mass regime, with $\leq$20 epochs of observations. Higher resolution and greater instrument stability indicate that photon limited precisions of 2 \ms\ are attainable on moderately rotating M dwarfs (with \vsini\ $\leq$5 \kms) using our technique.
\end{abstract}


\section{Introduction}

It is reasonable to expect that planets with predominantly lower mass may be found orbiting low mass stars. Core accretion theory predicts that the formation of Jupiter mass planets orbiting M dwarf stars is seriously inhibited \citep{laughlin04}. This argument is in good agreement with the Keck survey observations which estimate that the fraction of Jupiter-mass planets orbiting M, K and G stars is 1.8\%, 4.2\% and 8.9\% respectively \citep{johnson07}. More recently, it has been shown that the occurrence rate of gas giants orbits of less than 2000 days is 3\,-\,10 times smaller for M dwarfs than for F, G and K dwarfs \citep{cumming08}. In fact, models predict even lower incidences of Jupiter-mass planets, although the trend of a decrease in numbers with decreasing stellar mass agrees with observations. For instance, \cite{ida05} predict $\sim$\,7.5 times lower incidence of hot Jupiters around 0.4 \msun\  when compared with 1.0 \msun\ and \cite{kennedy08} predict a factor of $\sim$\,6 difference.

However, for lower mass protoplanetary disks, predictions by \cite{ida05} indicate that although ice cores can form, they can still acquire mass through inefficient gas accretion. The resulting bodies form Neptune mass (or $\sim\,$ 10 \mearth) planets which are able to undergo type II migration, ending in close proximity to the parent star, with radii $<$ 0.05 AU. It has also been noted that the semi-major axis distribution of known radial velocity planets orbiting M stars peaks at closer orbits than for more massive stars \citep{currie09}. While it is expected that the frequency of Neptune mass close-in planets peaks for M $\sim\,$ 0.4 \msun\  stars, the mass distribution peak is expected at 10 \mearth\ for 0.2 \msun\  stars, with a tail extending down to a few \mearth. In fact the Kepler planetary candidates now confirm that \mearth to M$_{Nept}$\ planets are $\sim$\,2.5\,-\,4 times more numerous around M stars than earlier spectral types \citep{borucki11kepler}, and crucially are found in greater numbers 
than transiting Jupiter size planets orbiting earlier spectral types. Many of these candidates are in very close orbits of $\lesssim$ 0.1 AU. Moreover, there are indications \citep{howard10,wittenmyer11} that 11.8\%-\,17.4\% of solar-like stars possess rocky (1\,-\,10 M$_{\rm Earth}$) planets. 
Scaling these values by the 2.5\,-\,4 times increased rocky planet occurrence rate reported by Kepler indicates that {30\,-\,70 per cent of M dwarfs should be orbited by close-in rocky planets. \cite{bonfils11mdwarfs} have recently reported on occurrence rates from the early-M dwarf {\sc harps} (High Accuracy Radial Velocity Planet Searcher) sample and find a similarly high occurrence rate for rocky planets ($\eta_\oplus$) of 0.54$^{+54}_{-13}$.}
Similarly, for Neptune mass planets, scaling the respective \citep{howard10,wittenmyer11} frequencies of 6.5 and 8.9 per cent by the Kepler candidate planet frequencies leads to expected occurrence rates of 13\,-\,27 per cent for M dwarfs. {Whether {\em rocky} planets that orbit in the classical habitable zone (where liquid water could be present) are in reality habitable, may depend on factors such as tidal locking and stellar activity \citep{kasting93hz,tarter07}. More recent studies, that investigate factors such as obliquity may also be important considerations for habitability \citep{heller11tidal}, although these topics are beyond the scope of this paper which focusses merely on planetary detection.}

Despite comprising 70\% of the solar neighbourhood population, the lower mass M dwarfs have remained beyond efficient detection with optical based surveys as the stellar flux peaks at longer wavelengths. At infrared wavelengths, surveys targeted at searching for low mass planetary systems associated with low mass stars can be achieved with realistic timescales. Nevertheless, to date, only one survey, using CRIRES \citep{bean10b}, has reported sub-10 \ms\ precision, with 5.4 \ms\ reported using an NH$_3$ gas cell and modest 364 \AA\ of spectrum in the K-band. More recent surveys have used telluric lines as a reference fiducial, and while \cite{rodler12keck} have obtained 180\,-\,300 \ms\ precision at R $\sim$20,0000, \cite{bailey12keck} have achieved $\sim$50 \ms\ precision which is limited by activity on their sample of young active stars. Equally, the red-optical regime offers similar advantages with a V\,-\,I $\sim$\,2\,-\,5 for the earliest-latest M dwarfs respectively, {but with much improved wavelength coverage over the near-infrared regime of \cite{bean10b}}. There are also significant telluric free spectral windows in the red-optical. 

The effects of jitter due to stellar activity are expected to present limitations but are also expected to be lower for M dwarfs than for F, G \& K dwarfs since the contrast ratio between spots and stellar photosphere is lower and the starspot distributions may be more uniformly distributed across the stellar surface \citep{barnes01mdwarfs}. We have simulated the radial velocity (RV) “noise” for M dwarfs (using a 3D, 2-temperature stellar model) with low (solar) activity levels and estimate that the radial velocity jitter is $\sim$\,1.5\,-\,2 times smaller in the I-band vs the V-band \citep{barnes11jitter}. We estimate that the appropriate jitter in the near infrared Y-band for even an extreme solar activity M dwarf analogue with \vsini\ = 5 \kms\ is in the range $\sim$0.7\,-\,5 \ms\ for an M6V star. The uncertainty arises from estimates of extremes of spot contrast, with 0.7 \ms\ corresponding to a photosphere-spot temperature difference of T$_{\rm phot}$ - T$_{\rm spot}$ = 200 K, while 5 \ms\ arises when the contrast is much higher, with T$_{\rm spot}$/T$_{\rm phot}$ = 0.65. Given that mean solar spot temperatures are of order  T$_{\rm phot}$ - T$_{\rm spot}$ = 500 K \citep{lagrange10spots}, we can assume that $\sim$2\,-\,3 \ms\ is a reasonable estimate of the spot induced jitter of a 5 \kms\ star when we scale our results to the I-band.

Flaring is known to be an important contributor to variations in spectral features in stars, but being high energy phenomena tend to affect bluer wavelengths more than redder and infrared wavelengths. \cite{reiners09flare} investigated flaring in the nearby active M6 dwarf, CN Leo, from nearly 200 observations spanning three nights. The spectra, taken with {uves} at the Very Large Telescope array were taken at at R = 40,000, covering 0.64\,-\,1.08 \micron, and thus closely resemble our observation setup with {\sc mike} (\S \ref{section:observations}). It was found that a large flaring event, easily identifiable by observing strong transient emission in \ha, could effect radial velocity measurements by $\sim$500 \ms\ in orders containing strong emission features. After the flaring event, the precision was limited to \hbox{$\sim$50 \ms}. In other orders however, it was found that the radial velocity jitter is below the 10 \ms\ level, even during flaring events. Careful choice of line regions is therefore important and inspection of activity indicators, such as \ha\ important where sub-10 \ms\ precision is required.

In section \S \ref{section:observations} we discuss our observations and data reduction and demonstrate the case for attempting precision radial velocities of M dwarfs at red-optical wavelengths. In \S \ref{section:wavelength}, we justify our choice of reference fiducial and give details of our wavelength calibration procedure. \S \ref{section:method} introduces the least squares deconvolution approach to measuring precision radial velocities before we present results from simulations and our observations in \S \ref{section:results}. In light of our observational radial velocities, a discussion (\S \ref{section:discussion}) that further considers the sources of noise in our results is given, before we make make concluding statements of the prospects of this kind of survey (\S \ref{section:conclusion}).

\begin{figure}
\begin{center}
\includegraphics[height=8.0cm,width=5.cm,angle=270]{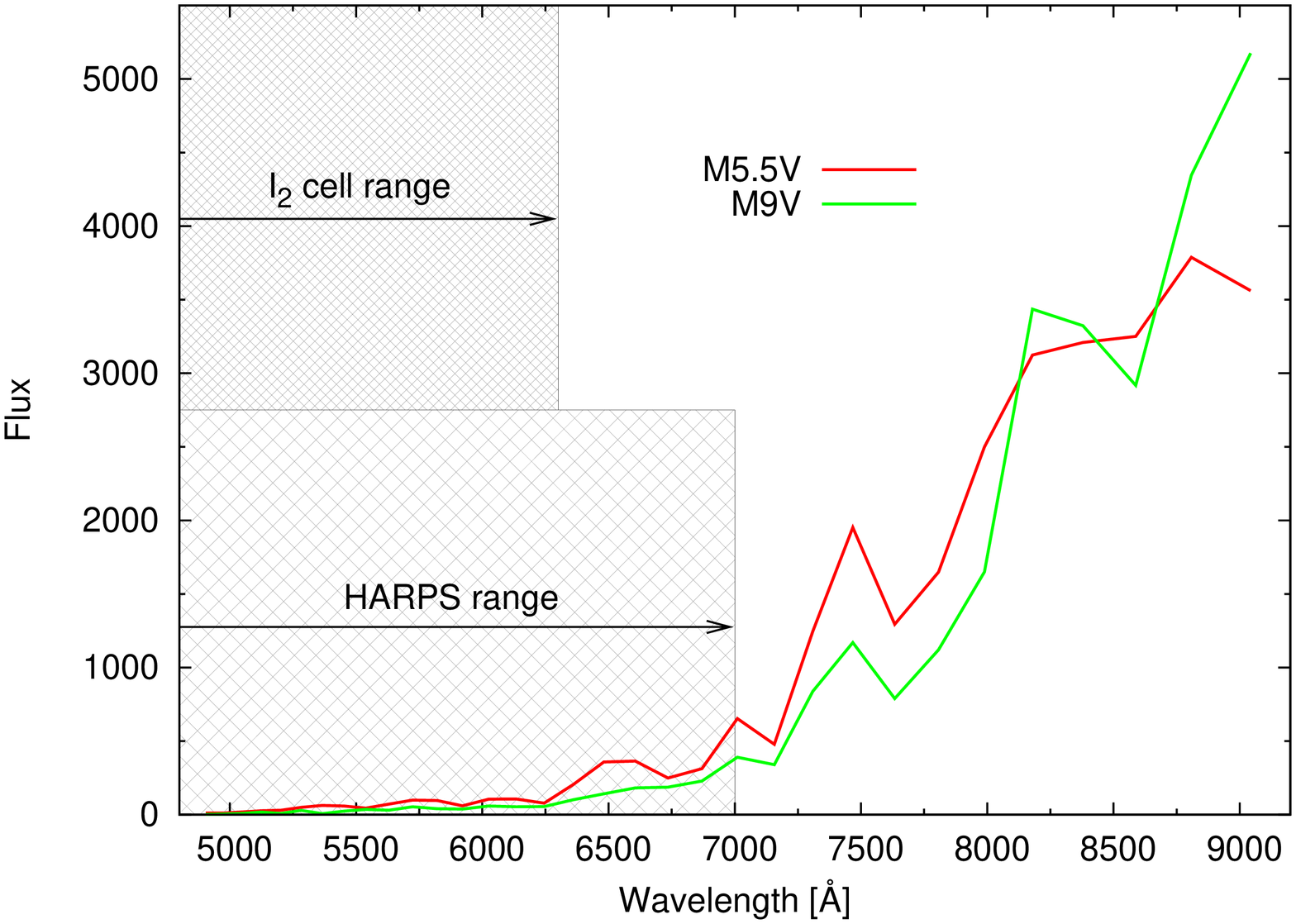} \\
\end{center}   
\caption{The mean flux level of the 33 extracted {\sc mike} orders covering 4830\,-\,9172 \AA\ for M5.5V and M9V targets (scaled to the same level at I band centre). Also shown are the optical I$_2$ cell regime and the HARPS coverage that overlap with the {\sc mike} red arm data. We find that F$_{0.7-0.9}$/F$_{0.5-0.7}$ = 11.5 and 19 for the M5.5V and M9V spectra respectively, demonstrating the inefficiency of wavelengths $<$ 7000 \AA for observing M dwarf targets.}
\protect\label{fig:fluxes}
\end{figure} 

\section{Observations}
\protect\label{section:observations}

We observed a number of bright M dwarf targets with the Magellan Inamori Kyocera Echelle ({\sc mike}) spectrograph at the 6.5 m Magellan Clay telescope on 2010 November 21 \& 22. Although the instrument simultaneously records two wavelength regions in a blue arm 3350\,-\,5000 \AA\ and a red arm 4900\,-\,9500 \AA, we only made use of the latter owing to the low S/N ratio obtained on our mid-late M dwarf targets. We observed with a 0.7\arcsec\ slit.

{Since ThAr lamps exhibit many lines for calibration, and are generally always used with \'{e}chelle spectrographs working at optical wavelengths, we made regular observations with the comparison lamp available with {\sc mike}. {\sc harps} spectra have been used by \cite{lovis07thar} to determine ThAr line wavelengths with a precision of $\sim$ 10 \ms (0.18 m\AA) on average for individual lines. This enables global wavelength solutions that attain sub-\ms\ stability when several hundred lines, spanning many \'{e}chelle orders, are used.} The median FWHM of the ThAr lines was found to be 4.05 pixels, indicating a mean resolution of R = 34,850. The observing conditions over the two nights were very good, with seeing estimates in the range 0.7\,-\,1.2 for targets observed at airmasses $<$ 1.5. Our targets are listed in Table \ref{tab:targets}, and with the exception of LP944-20 are all known to exhibit \vsini\ $\leq$ 5 \kms (Jenkins et al., In prep). Our observing sequence comprised of taking spectra alternated with calibration frames since it is not possible obtain a simultaneous reference observation with {\sc mike} at red wavelengths. We note that {\sc mike} does possess an iodine cell but that the wavelength cutoff of $\sim$ 7000 \AA\ means that we are unable to make use of I$_2$ lines.

\subsection{Data extraction}
Pixel to pixel variations were corrected for each frame using flat-field exposures taken with an internal tungsten reference lamp. Since few counts are recorded in the reddest orders of the red CCD on MIKE when care is taken to keep counts in the middle orders to no more than 50,000 counts (full well capacity 65,536 counts), we resorted to taking two sets of flat field frames of different exposure lengths. A total of 60 exposures of 40s each were used to flatfield the reddest orders. The worst cosmic ray events were removed at the pre-extraction stage using the Starlink {\sc figaro} routine {\sc bclean} \citep{shortridge93figaro}. The spectra were extracted using {\sc echomop}'s implementation of the optimal extraction algorithm developed by \citet{horne86extopt}. {\sc echomop} rejects all but the strongest sky lines \citep{barnes07b} and propagates error information based on photon statistics and readout noise throughout the extraction process. 

\subsection{M dwarf spectral fluxes at red-optical wavelengths}
\protect\label{section:fluxes}

Fig. \ref{fig:fluxes} compares the wavelength regimes used to make precision radial velocities. We note that by utilising the 0.62\,-\,0.90 \micron\ region, we use almost the same wavelength extent as {\sc harps} which covers 0.38\,-\,0.69 \micron. The fluxes observed with {\sc mike} are shown for an M5.5V spectrum and an M9V spectrum. The spectral energy distributions are determined from the observations with no flux calibration, and thus include the spectral energy distribution of the stars and the response of the telescope and instrument (i.e. CCD quantum efficiency and instrumental efficiency and blaze function). The curves are derived by taking the mean counts in each spectral order recorded on the red arm CCD of MIKE. The benefit of using red-optical wavelengths for mid-late M dwarfs is obvious with a recorded flux in the 0.7\,-\,0.9 \micron\ regime that exceeds the flux measured at 0.5\,-\,0.7 \micron\ (traditionally used by surveys that utilise ThAr or iodine reference lines), by a factor of 11.5 (M5.5V) to 19 (M9V). This is equivalent to a magnitude advantage of $\sim$2.65\,-\,3.2. 

{ The Doppler information available in these regimes is equally important and must be taken into consideration when estimating velocity precisions achievable in different wavelength regimes, as found by \cite{reiners10rvs} for example. For a M5.5V dwarf, the total number of lines available to {\sc harps} is $\sim$ 16,000, while the 0.6\,-\,0.9 \micron\ region contains \hbox{$\sim$ 12,300} lines that are on average 14 per cent less deep \citep{brott05}. Similarly, the number of lines in the \hbox{0.5\,-\,0.7 \micron} region plotted in Fig. \ref{fig:fluxes} is \hbox{$\sim$10,3000}, while only \hbox{$\sim$ 7400} lines are available in the 0.6\,-\,0.9 \micron\ region, with relative normalised depths of 0.67. For an M5.5V dwarf, assuming Poisson statistics correlate linearly with precision, the relative precision for a fixed integration time in the red-optical is expected to be 2.4 times that attained at standard optical wavelengths with the same integration time. In other words, to obtain the same precision at standard optical wavelengths would require 5.6 times the integration time. Similarly, for an M9V dwarf, the models of \cite{brott05} indicate line number and depth ratios (0.7\,-\,0.9 \micron\ / 0.5\,-\,0.7 \micron) of 0.755 and 0.925 respectively. Precision increases by 3.6 times when working in the red-optical, or equivalently integration times of only $\sim$1/13 are required to achieve the same precision as standard optical wavelength surveys.} 

\begin{table*}
\begin{tabular}{lccccccc}
\hline
Star     		& SpT   & Imag & Exp    & $v$ sin $i$ & S/N (extracted) & S/N (deconvolved) & Number of observations \\
                        &       &      & [s]    & kms$^{-1}$ &            &                   &                        \\
\hline
GJ 1002   		& M5.5V & 10.2 & 300    &   $\leq$5 & 113 & 4610   & 8  \\
GJ 1061   		& M5.5V & 9.5  & 300    &   $\leq$5 & 163 & 7400   & 12  \\
GJ 1286   		& M5.5V & 11.1 & 350    &   $\leq$5 &  84 & 3480   & 8  \\
GJ 3128   		& M6V   & 11.1 & 350    &   $\leq$5 &  84 & 3480   & 8  \\
SO J025300.5+165258	& M7V	& 10.7 & 350	&   $\leq$5 &  90 & 3150   & 9  \\ 
LHS 132			& M8V 	& 13.8 & 500	&   $\leq$5 &  25 & 875    & 4  \\
LP944-20 		& M9V   & 13.3 & 500    &  30       &  44 & 1540   & 7  \\
\hline
\end{tabular}
\vskip 2mm
\caption{List of targets observed on November 21 \& 22 with their estimated spectral types, I band magnitudes, \vsini\ values (Jenkins et al., In prep.) and exposure times. Extracted S/N ratio and S/N ratio after deconvolution are tabulated in columns 6 \& 7. Column 8 lists the total number of observations on each target.}
\protect\label{tab:targets}
\end{table*}

\section{Wavelength calibration}
\protect\label{section:wavelength}

Wavelength calibrations were made by an initial order-by-order identification of suitable lines using the ThAr wavelengths published by \cite{lovis07thar}, which are estimated to enable a calibration {(i.e. global)} r.m.s to better than { 20 \cms}\ for {\sc harps} {(see \S \ref{section:observations})}. Pixel positions were initially identified for a single arc using a simple Gaussian fit. For each subsequent arc, a cross-match was made followed by a multiple-Gaussian (up to three profiles) fit around each identified line using a Levenberg-Marquardt fitting algorithm \citep{press86} to obtain the pixel position of each line centre. The Lovis \& Pepe line list was optimised for {\sc harps} at R = 110,000, while our observations were made at R $\sim$ 35,000 necessitating rejection of some lines. Using a multiple Gaussian fit enables the effect of any nearby lines to be accounted for in the fit that also included a first order (straight line) background. Any lines closer than the instrumental FWHM were not used. Finally for each order, any remaining outliers were removed after fitting a cubic-polynomial. In addition, any lines that were not consistently yielding a good fit for all arc frames throughout both nights (to within 3-sigma of the cubic fits) were removed. A total of 32\,-\,71 lines were available for the orders centered at 6363\,-\,7470 \AA. For the redder orders centered at 7636\,-\,8811 \AA, 18\,-\,25 lines were available. To improve stability, we made a two dimensional wavelength calibration, with 4 coefficients in the dispersion direction and 7 coefficients in the cross-dispersion direction. By iteratively rejecting outlying pixels from the fit, we found that of the input 575 lines, clipping the 15 furthest outliers yielded the most consistent fit from one solution to the next. Sigma clipping has the potential to remove blocks of lines in a given region and we found this to give less stable solutions. Of the 575 input lines, we clipped 15 from each frame leaving a total of 550 lines for each solution. The zero point r.m.s. (i.e. the r.m.s. by combining all lines) is found to be 8.73 $\pm$ 0.32 \ms. { Wavelength calibration precision could be improved, in the reddest orders at least, where $\lambda >$ 0.85 \micron, by simultaneously using a ThAr and UNe lamps. The UNe lamp has been shown to provide around 6 times more lines in this region than the ThAr lamp \citep{redman11uranium}}.

We find that despite a precision of sub 10 \ms\ on a given wavelength solution, the solution from one ThAr frame to the next yields r.m.s. residuals of 30\,-\,80 \ms. The difference does not show a constant shift or tilt across all orders but may appear random from one order to the next indicating that the long term solution is in fact not stable. The dynamic range of the ThAr lines used was however large, so that weak lines, especially in the redder order where fewer lines per order are available, contribute to a less stable solution that is desirable. A total of 88 per cent of the lines possess strengths that are $\leq$ 0.05 of the maximum line used (even considering the removal of lines that peaked with more than 50000 counts before extraction), with most of these lines possessing only a few thousand counts above the bias level.

\subsection{Stability of {\sc mike}}

\begin{figure}
\begin{center}
\vspace{10mm}
\includegraphics[angle=270,bbllx=104,bblly=138,bburx=496,bbury=654,width=62mm]{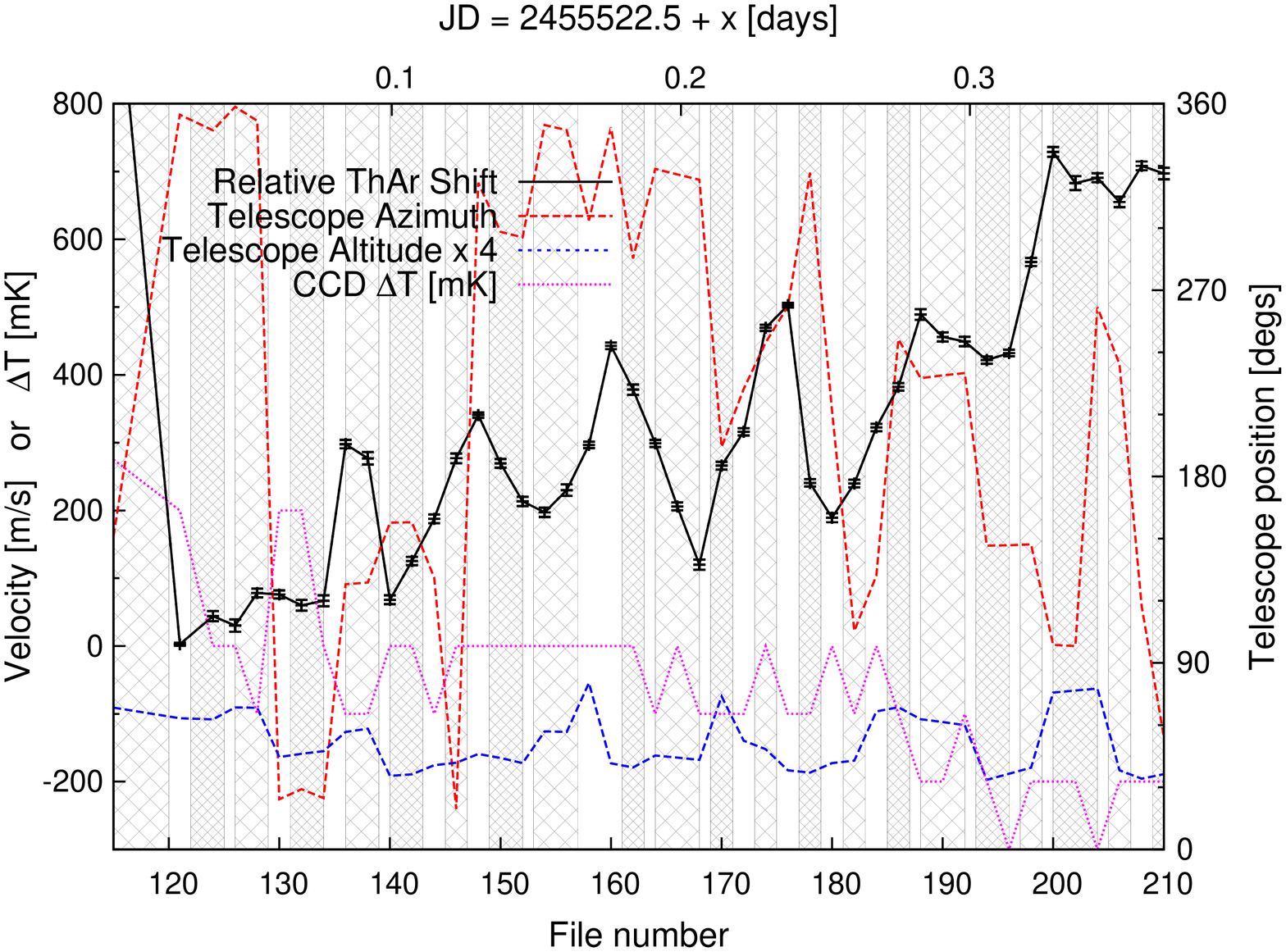} \\
\vspace{18mm}
\includegraphics[angle=270,bbllx=104,bblly=149,bburx=496,bbury=654,width=62mm]{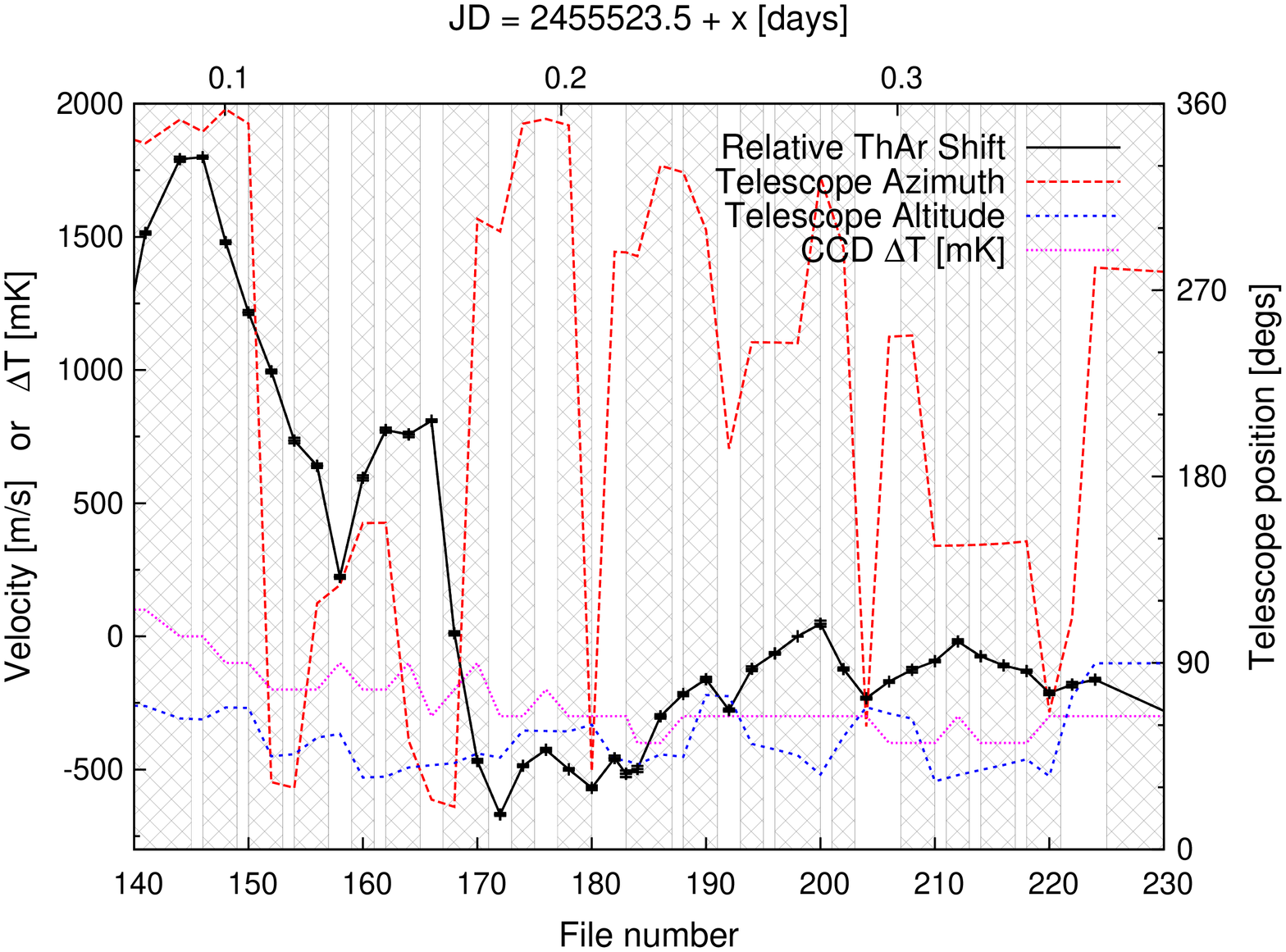} \\
\end{center}   
\vspace{5mm} 

\caption{Stability of {\sc mike} on the 21st (top) and 22nd (bottom) measured using cross-correlation of ThAr lines (crosses joined by solid line). Velocity in ms$^{-1}$ is plotted against observed frame sequence number. Also plotted are the corresponding temperature variation of the dewar and the position (as altitude and azimuth) of the telescope. The temperature is measured in K and multiplied by 1000, while the azimuth can be read directly in degrees. The telescope altitude is multiplied by 4 for clarity. The hatched regions indicated periods over which the telescope was pointing at a fixed RA and Dec on the sky (i.e. tracking), and over which no slewing of the telescope took place.}
\protect\label{fig:stabilityplot}
\end{figure}

In Fig. \ref{fig:stabilityplot}, we illustrate the stability of {\sc mike} by cross-correlating ThAr frames. This enables us to ascertain our precision when interpolating our reference velocity for each target observation. The {\sc mike} User's Guide\footnote{http://www.ucolick.org/$\sim$rab/{\sc mike}/usersguide.html} indicates that the instrument is stable at the $\sim$ 1 pixel level (in the cross-dispersion direction), owing to small temperature changes through the night. However, Fig. \ref{fig:stabilityplot} in fact indicates that use of {\sc mike} as a precision radial velocity instrument requires that a simultaneous reference fiducial be used if consistent results are to be achieved. The crosses joined by solid lines represent the drift in ms$^{-1}$ as a function of frame number (covering all science frames taken throughout the night). The hatched regions represent times at which the telescope was pointing at the same right ascension and declination; in other words, no slewing occurred during these 
intervals. Nevertheless, it can be seen that large shifts of the instrument take place during these intervals. In some cases, particularly on the second night, the wavelength appears to have drifted by $\sim$ 500 ms$^{-1}$, or $\sim$ 0.25 pixels between arc observations. In many cases, the drift is somewhat less, but clearly severely prohibits our ability to make precision radial velocity measurements of $\leq$ 10 ms$^{-1}$ by relying on ThAr measurements. 

Moreover, inspection of telluric lines indicates changes in position that do no correlate with the arc lines. We suspect that this apparent discrepancy, and the semi-random nature of the measured offsets may arise from mechanical movement of the calibration mirror when alternating between ThAr observations and science target observations.

\subsection{Precision velocity measurements with {\sc mike}}

\begin{figure}
\begin{center}
\includegraphics[angle=270,bbllx=104,bblly=149,bburx=496,bbury=654,width=62mm]{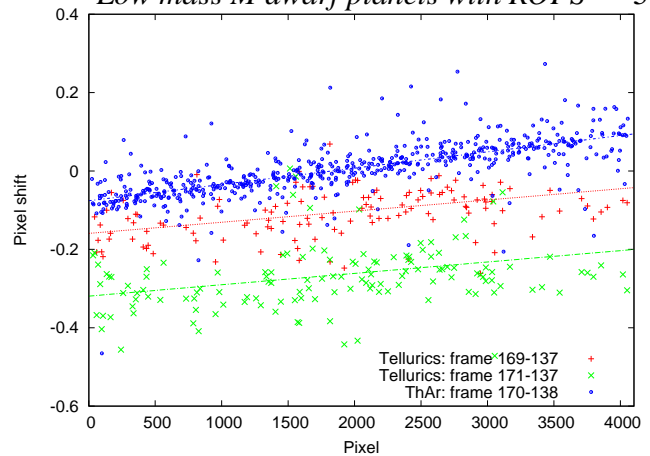} \\
\end{center}   
\vspace{5mm} 

\caption{Comparison of Telluric shifts and ThAr shifts. The telluric lines positions were obtained from the second pair of observations of GJ 1061 (our target with the highest S/N) on the first night. The shifts are plotted relative to the first observation of GJ 1061 as plus symbols (red) and crosses (green). Similarly the shift in the ThAr positions is plotted for the ThAr frame taken between the GJ 1061 observation pair relative to the first ThAr frame (taken after the first GJ 1061 frame) and are plotted as small points (blue).}
\protect\label{fig:stabilityplot2}
\end{figure}

While Fig. \ref{fig:stabilityplot} indicates that the ThAr lines can shift by large quantities, Fig. \ref{fig:stabilityplot2} demonstrates the relationship between the ThAr shifts and the shifts from telluric lines (i.e. the telluric lines do not show the same shifts as the ThAr lines). The plot shows the relative drift in pixels as a function of pixel for all orders, clearly indicating that a tilt is also present. We measured the strongest telluric lines from observations of GJ1061 since this object has the highest S/N \hbox{$\sim$ 160}. We measured the telluric line positions in the same manner as described above for the ThAr frames. Plotted in Fig. \ref{fig:stabilityplot2} are the shifts relative to the first observation of GJ 1061 on the first night. It can be seen that there is a significant shift, even between the pair of observations, which were made together with a ThAr observation between them. Similarly, we compared the ThAr frame between these two observations with that taken immediately after 
the first GJ 1061 observation. The scatter in the telluric lines is too great to give a reliable tilt measurement, but the ThAr relationship is much tighter. Nevertheless, the tilt appears to be approximately the same for the tellurics and the ThAr frames, even of the shift is different. The similarity of tilt, but differing shift is seen for other pairs of observations of GJ 1061. However, owing to the larger scatter seen in telluric pixel positions for all other objects (which typically possess 0.5\,-\,0.75 of the S/N of our GJ1061 observations), we are unable to easily see the tilt. 

Our strategy for making precision radial velocities stems from the above observations. We assume that the local wavelength solution can be described as a shift and a tilt relative to a single master spectrum. Since we find that the local 2D wavelength solution results in variations of 30\,-\,80 \ms (\S \ref{section:wavelength} above), we instead use the relative tilt in each wavelength solution to correct the master wavelength frame (chosen as the wavelength solution at the start of observations for each target star). Since the tilt is derived by combining all ThAr orders, we then apply this uniformly to each order in turn for the relevant ThAr observation that is made between each observation pair. This procedure is effectively equivalent to allowing only the low order terms to vary in the 2D wavelength solution. Each wavelength corrected frame is then used to make the subsequent deconvolved profiles (see next \S \ref{lab:lsd}) for the pair of observations that bracket it. The local shift for a given observation is made by measuring the {\em shift} of the telluric lines since this shift is not the same as that seen in the ThAr frames. Since this latter shift is assumed to apply to both the stellar and telluric lines, it is limited by the stability of the telluric lines. However, the correction, even for significant shifts only affects the zero point wavelength in our calculations. For example, even a shift of 1 pixel (equivalent to 2124 \ms) at 8000 \AA\ modifies the zero point wavelength by $\Delta\lambda = \lambda_0\Delta v / c$ = 0.057 \AA. The corresponding velocity error arising from a 0.057 \AA\ shift error ($\lambda_0 = 8000 \pm 0.057$) is only \hbox{$\sim$0.015 \ms}.

\section{Precision Radial velocities of M dwarf spectra}
\protect\label{section:method}

{Since we are unable to achieve the desired sensitivity for precision radial velocities using the internal ThAr emission line source, an alternative simultaneous fiducial is needed. To date, the only {\em gas cell} to be identified for precision work at optical wavelengths uses iodine (I$_2$) as a reference fiducial \citep{marcy92iodine} as it possesses a rich source of absorption lines. However, as Fig. \ref{fig:fluxes} illustrates, the I$_2$ lines do not extend to the wavelengths we are using.} In the near infrared, there has been some effort made towards identifying suitable cells (e.g. \citealt{mahadevan09gascells}), while the only successful infrared gas cell survey has utilised a NH$_3$ gas cell that provides lines in the K band \citep{bean10b}. The use of a laser comb to provide regularly spaced reference lines \citep{steinmetz08} has now demonstrated 10 \ms\ on stellar targets \citep{ycas12lasercomb}, although this technology is still being developed (at infrared wavelengths) and currently remains an expensive choice. The only available option for our data is to investigate the possibility of making use of the numerous telluric reference lines available in our spectra. Hence the intuitive requirement of stellar spectral regions that are void of atmospheric lines is replaced by a strategy that makes full use of the recorded spectral range in each exposure. We are therefore faced with a situation where the observed spectrum and reference lines are often superimposed, in much the same manner as {\em when using a gas cell inserted into the instrument light path.} Our adopted method is dictated by the nature of the spectra in the sense that we do not possess a reliable stellar template spectrum that is free of the reference spectrum (the telluric lines in this case). While accurate reference spectra can be derived from atmospheric models, the stellar spectra, at least at high resolution, are not highly accurate, necessitating 
a semi-empirical approach. Rather than cross-correlating small regions of spectra and optimally co-adding, we employ a hybrid of the iodine technique of Marcy et al. and the ThAr method first described by \cite{baranne96} and now incorporated into the HARPS pipeline. 

\subsection{Least Squares Deconvolution}
\protect\label{lab:lsd}

We have applied least squares deconvolution (LSD) to our spectra to derive a single high S/N line profile for each observed spectrum. In order to disentangle the stellar signature from the telluric signature, we apply the technique using models that represent the stellar lines and the telluric lines individually i.e. a stellar line list and telluric line list giving line positions and normalised line depths. Hence, two simultaneous high S/N ratio line profiles are derived, one from the stellar lines and one from the telluric lines. The deconvolved profiles are derived in velocity space, enabling direct relative shifts to be measured.

The deconvolution procedure has been used in a number of applications but was first implemented by \cite{donati97zdi} to boost the S/N in weak Stokes V profiles in noisy, high resolution timeseries spectra, thereby enabling indirect stellar magnetic images to be derived. It was also implemented by \cite{barnes98aper} to similarly enable inversion of intensity profiles to produce high resolution surface brightness maps of relatively faint, rapidly rotating Alpha Persei G dwarfs. More recently, the procedure has been used to search for the weak signature of planetary signals in high resolution optical \citep{cameron99tauboo,cameron02upsand,leigh03b} and infrared \citep{barnes07a,barnes07b,barnes08,barnes10hd189733b} spectra. LSD is also effective for deriving accurate \vsini\ values for low S/N spectra that do not necessarily contain strong photospheric lines. The \vsini\ values of a number of M dwarfs were derived in this way in \cite{jenkins09mdwarfs}. We employ the same technique in Jenkins et al. (in prep.) to derive the \vsini\ values for our sample of M dwarfs in this paper.

Although the method has been described before, here we give a detailed account of our implementation to emphasize its application to precision radial velocity measurements.
If the deconvolved profile elements are denoted by {\mbox{\boldmath$z_k$}}, then the predicted data, {\mbox{\boldmath$p_j$}} (i.e. the convolution of the line list and deconvolved profile), can be written as

\begin{equation}
{\mbox{\boldmath${p_{j}}$}} = \sum_k{{\mbox{\boldmath{\sf{$\alpha$}}}}}{\mbox{\boldmath{$_{jk}z{_k}$}}},
\protect\label{eqn:masktimeslsdprof}
\end{equation}
\noindent
The elements of the $N_{j} \times N_{k}$ matrix, {\mbox{\boldmath$\alpha$}}, are defined by the line depths, {\mbox{\boldmath$d_i$}}, and the triangular interpolation function, $\Lambda(x)$, as

\begin{equation}
{\mbox{\boldmath$\alpha_{jk}$}} = \sum_i {\mbox{\boldmath$d_i$}} \Lambda(x)
\protect\label{eqn:designmatrix}
\end{equation}

\noindent
where 

\noindent
\begin{equation}
x = \left({\mbox{\boldmath$v_k$}}-c\left(\frac{{\mbox{\boldmath$\lambda_j$}}-{\mbox{\boldmath$\lambda_i$}}}{{\mbox{\boldmath$\lambda_i$}}}\right)\right)/\Delta v, \\
\protect\label{eqn:define_x}
\end{equation}

\noindent
and \\
{\mbox{\boldmath$d_i$}} = {\rm depth of line} {\em i} \\
{\mbox{\boldmath$\lambda_i$}} = {\rm wavelength of line} {\em i} \\
{\mbox{\boldmath$\lambda_j$}} = {\rm wavelength of spectrum pixel} {\em j} \\
{\mbox{\boldmath$v_k$}} = {\rm radial velocity of profile bin} {\em k} \\
$\Delta v$ = {\rm velocity increment per pixel in deconvolved profile} \\

\noindent 
The triangular function $\Lambda(x)$ is such that

\begin{equation}
\begin{tabular}{llll}
$\Lambda(x)$ & = $1 + x$ & {\rm for} & $-1 < x < 0$ \\
             & = $1 - x$ & {\rm for} & $~0 < x < 1$ \\
	     & = 0       & \multicolumn{2}{l}{elsewhere}  \\  
\end{tabular}
\protect\label{eqn:designmatrix_lambda}
\end{equation}

\noindent
In this way, the weight of each spectrum element, $j$, is partitioned across 2 or more pixels, $k$. This interpolation in velocity space from each observed line to the deconvolved line thus takes account of the change in dispersion seen across the spectral orders. Hence a sparse matrix {\mbox{\boldmath$\alpha$}} is built up such that it contains minor diagonals, usually with pairs of partitioned elements in $k$ for each element or pixel in $j$. In other words, the method given in equations \ref{eqn:designmatrix}, \ref{eqn:define_x} and \ref{eqn:designmatrix_lambda} is defined such that the interpolated pairs of elements in $k$, at a given $j$, add to give the depth of line $i$ in question. The design matrix, {\mbox{\boldmath$\alpha$}}, is effectively a weighting mask which only includes the regions of spectrum over the desired deconvolution range in wavelength/velocity space. 

Simply convolving the normalised, observed spectrum (elements, {\mbox{\boldmath$r_j$}}) with the matrix, {\mbox{\boldmath$\alpha$}}, i.e.

\begin{equation}
{\mbox{\boldmath${x_{k}}$}} = \sum_j{{\mbox{\boldmath{\sf{$\alpha$}}}}}{\mbox{\boldmath{$_{jk}r{_j}$}}},
\protect\label{eqn:simple_crosscorr}
\end{equation}

\noindent
would be equivalent to co-adding all the weighted lines in the object spectrum which occur in the synthetic line list, but would not properly account for the effects of blended lines. Blending occurs in all spectra to some degree, being dependent on instrumental resolution and astrophysical properties, including stellar rotational line broadening. Blending occurs when there is more than one set of non-zero $k$ elements per $j$ element in the design matrix. The least squares deconvolution process means that the number of lines which can be used is independent of the degree of blending.

The output least squares profile is binned at the average pixel resolution of the object spectrum and is determined by the size of the CCD pixels (approx. 2.1\,-\,2.2 \kms\ in the case of our {\sc mike} observations). Generally the velocity range over which the deconvolution takes place is chosen to include continuum on either side of the profile. The \chisq~function is used to obtain the inverse variance weighted fit of the convolution of the line list with the deconvolved profile to the normalised and continuum-subtracted spectrum {\mbox{\boldmath$r_j$}},

\begin{equation}
\chi^2 = \sum_{j} \left( \frac{{\mbox{\boldmath{$r_{j}$}}}~-~\sum_{k}{\mbox{\boldmath{\sf{$\alpha$}}}}{\mbox{\boldmath{$_{jk}$}}}{\mbox{\boldmath{$z_{k}$}}}}{{\mbox{\boldmath$\sigma_j$}}} \right )^2.
\protect\label{eqn:lsd_equation}
\end{equation}
\noindent
The equation is minimised by finding the solution to the set of equations

\begin{equation}
\frac{\partial\chi^2}{\partial{{\mbox{\boldmath$z_{k}$}}}} = 0,
\protect\label{eqn:lsd_partial}
\end{equation}

\noindent
where the unknown quantity is the least squares profile, {\mbox{\boldmath$z_{k}$}}. This yields


\begin{equation}
\sum_j \frac{1}{{\mbox{\boldmath$\sigma_{j}^{2}$}}} \sum_k {\mbox{\boldmath{$\alpha_{jk}$}}}\sum_{l}{\mbox{\boldmath{\sf{$\alpha$}}}}{\mbox{\boldmath{$_{jl}$}}}{\mbox{\boldmath{$z_{l}$}}} = \sum_j \frac{1}{{\mbox{\boldmath$\sigma_{j}^{2}$}}}{\mbox{\boldmath{$r_{j}$}}} \sum_k {\mbox{\boldmath{\sf{$\alpha$}}}}{\mbox{\boldmath{$_{jk}$}}}
\protect\label{eqn:lsd_full}
\end{equation}

\noindent
The deconvolution process can be given in the same form as Equation 4 in \cite{donati97zdi}. If the inverse variances, 1/{\mbox{\boldmath $\sigma_j^2$}} are contained in the the vector, {\mbox{\boldmath$V$}}, and the spectrum elements {\mbox{\boldmath{$r_j$}}} in the vector, {\mbox{\boldmath$R$}}, the solution can be written in matrix form as

\begin{equation}
{\mbox{\boldmath$z$}} = ({\mbox{\boldmath$\alpha^T$}}.{\mbox{\boldmath$V$}}.{\mbox{\boldmath$\alpha$}})^{-1}~{\mbox{\boldmath$\alpha^T$}}.{\mbox{\boldmath$V$}}.{\mbox{\boldmath$R$}}
\protect\label{eqn:lsd_fullmatrix}
\end{equation}

The right hand part of Equation \ref{eqn:lsd_fullmatrix} (i.e. {\mbox{\boldmath$\alpha^T$}}.{\mbox{\boldmath$V$}}.{\mbox{\boldmath$R$}}) is the cross-correlation of the observed spectrum with the line mask. This, as mentioned above, is the weighted mean of all the lines. The solution however gives flat continuum outside the least squares deconvolved profile, free of side lobes from blends, provided that the line list used is reasonably comprehensive, and the weights correct. The solution therefore effectively cleans the cross-correlation vector from the square symmetrical auto-correlation matrix {\mbox{\boldmath$\alpha^T$}}.{\mbox{\boldmath$V$}}.{\mbox{\boldmath$\alpha$}}. The solution to equation \ref{eqn:lsd_fullmatrix} requires the inverse of the auto-correlation matrix to be determined. This can be found by making use of a suitable fast inversion routine such as Cholesky Decomposition as given by \cite{press86}. 


\begin{table*}
\begin{tabular}{ccccccccccc}
\hline
\multicolumn{3}{c}{GJ 1002}       & \hspace{12mm} & \multicolumn{3}{c}{GJ 1061}       & \hspace{12mm}  & \multicolumn{3}{c}{LHS 132}	 	\\ 
 HJD 	     & RV     &$\Delta$RV &              & HJD          & RV     &$\Delta$RV &               & HJD 	     & RV    &$\Delta$RV 	\\ 
 (2400000+)  & [\ms]  & [\ms]     &              & (2400000+)   & [\ms]  & [\ms]     &               & (2400000+)    & [\ms] & [\ms]     	\\            
\hline                                                                                                                             
~55522.538975&   8.18 & 11.47     & \hspace{12mm} &~55522.578698&  13.14  &  6.69    & \hspace{12mm}  & ~55522.602135 &  27.25 &  20.32 		\\
~55522.546660&  14.97 & 8.56      & \hspace{12mm} &~55522.585098&  30.05  &  3.88    & \hspace{12mm}  & ~55522.606359 &  20.27 &  32.18 		\\
~55522.666140&  38.07 & 11.97     & \hspace{12mm} &~55522.691451&  -0.94  &  6.04    & \hspace{12mm}  & ~55523.600360 &  -0.08 &  22.65 		\\
~55522.672159&  46.16 & 14.12     & \hspace{12mm} &~55522.697410&  -9.24  &  6.03    & \hspace{12mm}  & ~55523.605190 & -47.45 &  43.58 		\\
~55523.537970& -10.85 & 7.70      & \hspace{12mm} &~55522.760129& -17.53  & 16.32    & \hspace{12mm}  &  \multicolumn{3}{c}{\rule[0.75ex]{45mm}{0.4pt}} \\
~55523.545250& -33.82 & 14.95	  & \hspace{12mm} &~55522.766101&  -3.16  &  4.68    & \hspace{12mm}  &  \multicolumn{3}{c}{LP 944-20}        \\
~55523.658340& -39.21 & 16.76	  & \hspace{12mm} &~55522.852187& -13.86  & 11.21    & \hspace{12mm}  &  \multicolumn{3}{c}{\rule[0.75ex]{45mm}{0.4pt}} \\
~55523.665500& -26.14 & 17.24 	  & \hspace{12mm} &~55522.858298& -24.38  & 12.80    & \hspace{12mm}  & ~55522.579923 &  -41.88 &  36.58 		\\
	     &	      &	 	  & \hspace{12mm} &~55523.565640&   2.53  &  4.49    & \hspace{12mm}  & ~55522.622729 &  185.47 &  57.42 		\\
	     &	      &	 	  & \hspace{12mm} &~55523.574010&  -0.11  &  3.29    & \hspace{12mm}  & ~55522.626982 & -130.88 &  80.21 		\\
	     &	      &	 	  & \hspace{12mm} &~55523.687180&   2.70  & 14.34    & \hspace{12mm}  & ~55523.626580 &  -55.87 &  45.55 		\\
	     &	      &	 	  & \hspace{12mm} &~55523.694340&  20.80  & 22.70    & \hspace{12mm}  & ~55523.630760 &	43.17 & 125.84 		\\
\hline
\multicolumn{3}{c}{GJ 1286}       & \hspace{12mm} & \multicolumn{3}{c}{GJ 3128}&\hspace{12mm}& \multicolumn{3}{c}{SO J025300+165258}     		\\ 
 HJD 	     & RV     &$\Delta$RV &              & HJD          & RV      &$\Delta$RV &              & HJD 	     & RV     &$\Delta$RV	\\ 
 (2400000+)  & [\ms]  & [\ms]     &              & (2400000+)   & [\ms]   & [\ms]     &              & (2400000+)    & [\ms]  & [\ms]  		 \\            
\hline                                                                                                                                                                           
~55522.524989&  6.87 &  8.26  	  & \hspace{12mm}  &~55522.529716& -52.31  &  9.02     & \hspace{12mm} &  ~55522.607386 &  31.20 &  9.44		\\
~55522.532083& 24.11 &  9.56  	  & \hspace{12mm}  &~55522.533941& -43.36  &  7.72     & \hspace{12mm} &  ~55522.613983 &  30.96 &  7.03		\\
~55522.617712& 27.17 &  7.68  	  & \hspace{12mm}  &~55522.591953&  95.24  &  4.69     & \hspace{12mm} &  ~55522.708726 &  77.90 &  9.05		\\
~55522.624517& 19.41 &  7.27  	  & \hspace{12mm}  &~55522.595020&  109.71 &  8.51     & \hspace{12mm} &  ~55522.717175 &  89.68 &  9.95		\\
~55523.522780&-15.00 &  8.43  	  & \hspace{12mm}  &~55523.528760& -112.24 & 10.51     & \hspace{12mm} &  ~55523.593670 & -10.63 &  5.36		\\
~55523.529990&-13.07 &  8.98  	  & \hspace{12mm}  &~55523.531840&  -97.95 & 11.69     & \hspace{12mm} &  ~55523.601610 & -23.29 &  9.41		\\
~55523.614100&-20.65 &  8.64  	  & \hspace{12mm}  &~55523.588880&   29.41 & 11.00     & \hspace{12mm} &  ~55523.609830 & -23.52 &  6.81		\\
~55523.621280&-29.27 & 10.21  	  & \hspace{12mm}  &~55523.592470&   71.51 & 11.42     & \hspace{12mm} &  ~55523.727010 & -84.51 & 13.88		\\
	     &	     &	      	  & \hspace{12mm}  &   	         &     	   & 	       & \hspace{12mm} &  55523.734950 & -87.80 & 18.23		\\
\hline                                                                                                                                                                           
\end{tabular}
\caption{Radial velocities and cross-correlation uncertainties for each target.}
\protect\label{tab:rvs}
\end{table*} 

\subsection{Deconvolution line lists}
\protect\label{section:linelists}

For successful deconvolution, we require a good line list to represent both the synthetic telluric spectrum and the stellar spectrum. Since we find that spectra for mid-late M dwarfs \citep{brott05} do not adequately reproduce the line positions and strengths at the resolutions we are working at, we have adopted a semi-empirical approach for deriving line lists. We shifted and co-added spectra for one of our higher S/N targets (GJ 1002). Once a high S/N ratio template is obtained (S/N = 320 for all co-added \hbox{GJ 1002} frames), we simply search for absorption lines and find the line position by cubic interpolation of the line core. The line depths are measured at the same time for weighting in the deconvolution. We also included a S/N selection criterion to avoid selecting noise features as lines. Since we do not use lines that are weaker than 0.05 of the normalised depth, this should be avoided anyway since even lines with a depth of 0.05 are 16 sigma above the noise level. Finally, the lines in the stellar line list that were either close to known telluric lines (see below), or actual telluric lines, were rejected to remove the possibility of cross-contamination. The line list is not ideal since it is derived at the same resolution as the observation, whereas a higher resolution spectrum would be desirable to disentangle blends. Nevertheless, any incorrect wavelength positions from blended lines will be treated the same way in every deconvolution of a given target star.

To generate the synthetic telluric spectra, we use the Line By Line Radiative Transfer Model ({\sc lblrtm}) code \citep{clough92,clough05}, which is distributed by Atmospheric and Environmental Research ({\sc AER})\footnote{http://rtweb.aer.com}. The code uses the most up to date versions of the {\sc hitran} 2008 molecular spectroscopic database \citep{rothman09hitran}. We generate a reference spectrum at high resolution, using a meteorological sounding provided by Air Resources Laboratory ({\sc arl})\footnote{http://ready.arl.noaa.gov/READYamnet.php}. The soundings are available at a time resolution of 3 hours for the longitude and latitude of Las Campanas Observatory. The sounding contains an atmospheric observation containing wind speed, wind direction, temperature, dew point and pressure at a range of atmospheric heights and is used by {\sc lblrtm} to generate synthetic spectra for the current location and conditions. Our telluric line list is generated from a reference synthetic spectrum with a sounding taken on the first night of observations at Las Campanas observatory for an altitude of 60 degrees. Once a high resolution normalised spectrum is generated, we build a line list in the same manner as described above for the stellar lines, although contamination and S/N effects in this instance are not relevant.

\subsection{Derivation of radial velocities}
\protect\label{lab:rvs}

Radial velocity measurements are made by subtraction of the measured velocity centre of the deconvolved stellar profile with respect to the deconvolved telluric profile for each spectrum. We use a version of the {\sc hcross} algorithm of \cite{heavens93redshifts} which is in turn a modification of the \cite{tonry79} cross-correlation algorithm. {\sc hcross} applies a more accurate, robust algorithm that employs the theory of peaks in Gaussian noise to determine uncertainties in the cross-correlation peak. Since the routine, which is part of the Starlink package , {\sc figaro} (now distributed by the Joint Astronomy Centre\footnote{http://starlink.jach.hawaii.edu/starlink}), was originally intended for measurement of galaxy redshifts in low S/N spectra, we have made a minor modification to output both shift, and shift uncertainty, in pixels.

\begin{figure}
\begin{center}
\includegraphics[width=83mm,angle=0]{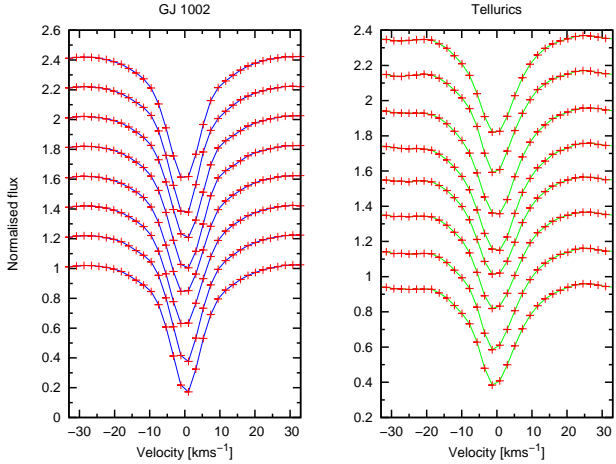} \\
\end{center}   
\caption{Right panel: Deconvolved profiles for GJ 1002 plotted in order of observation from top to bottom. The left panel shows the simultaneously deconvolved telluric profiles. Error bars are also plotted but too small to be visible.}
\protect\label{fig:profiles}
\end{figure} 

A master stellar line is derived by cross-correlating and aligning all deconvolved stellar lines. The master line is used for as a cross-correlation template since it more closely represents each individual line as opposed to using a simple Gaussian, Lorentzian or Voigt profile. A similar procedure is carried out for the telluric lines before the subtraction of pairs of stellar and telluric measurements can be made. Finally, we apply the barycentric corrections at the mid-time of each observation to correct our velocities to the solar-system centre. The final velocities are arbitrary and depend on the stellar line lists used (i.e. the radial velocity of GJ 1002). With our small sample spanning two days, we have opted to subtract the mean velocity from all data points of each target. With higher resolution template spectra, we will determine the absolute barycentric radial velocity of each observation in future work.

\section{Results}
\protect\label{section:results}

\subsection{Target star radial velocities}

\begin{figure*}
\begin{center}
\includegraphics[width=17.5cm,angle=0]{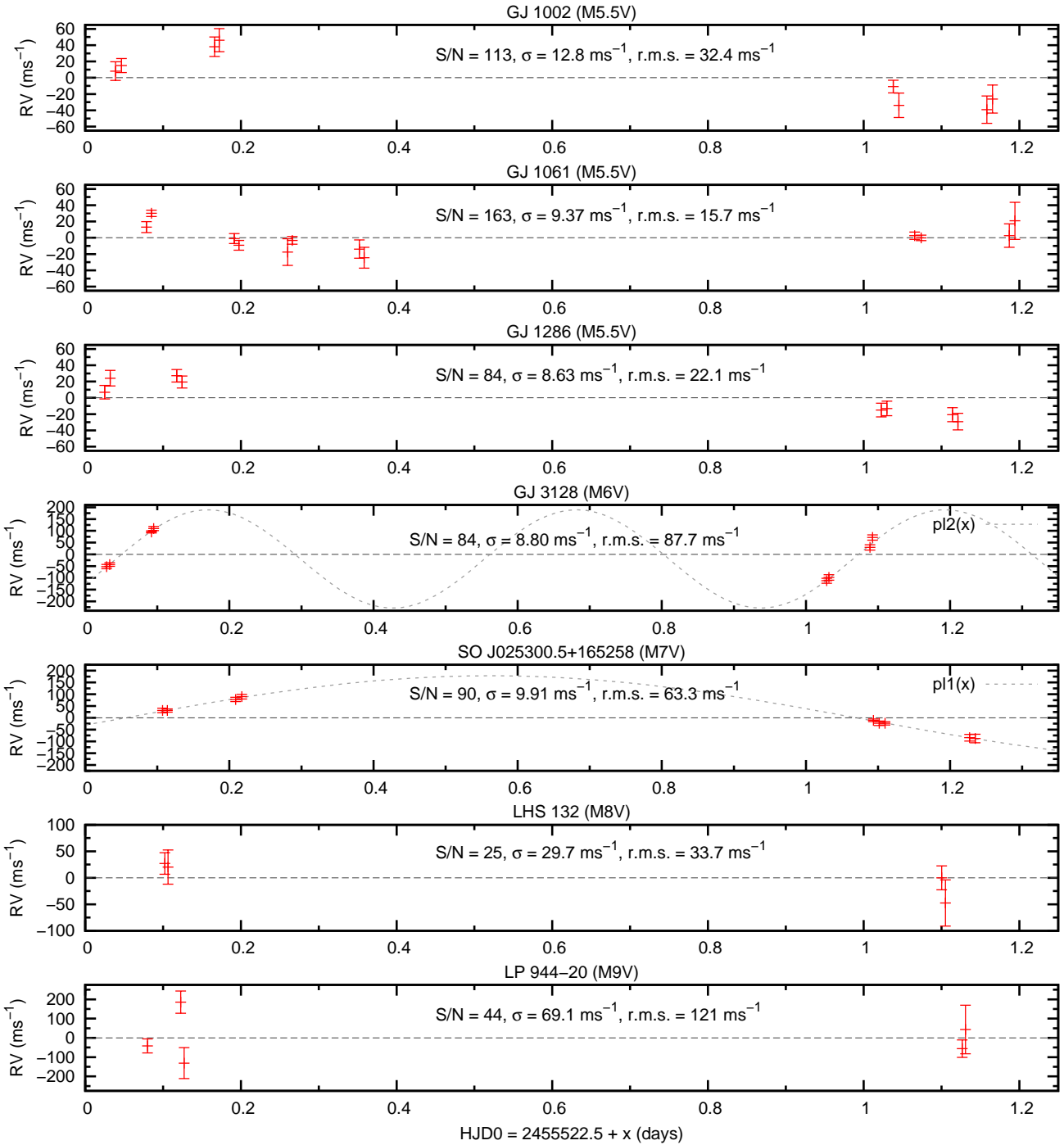} \\
\end{center}   
\caption{2010 November 21 \& 22 heliocentric corrected radial velocity measurements for the 7 M dwarfs, GJ 1002, GJ 1286 and GJ 1061, GJ 3128, SO J025300+165258, LHS 132 and LP944-20. The measurements were made by deconvolving the spectral lines in each frame into a single high S/N profile. The same procedure is applied to each telluric frame. The horizontal dashed line simply represents the mean velocity (zero). The pre-deconvolution S/N ratio, mean cross-correlation error, $\sigma$, and r.m.s. scatter in all points are given for each star. For GJ 3128 and SO J025300+165258, which both exhibit significant radial velocity variations, we fit a sinusoidal velocity curve (see \S 6.1 for discussion).}
\protect\label{fig:rvs}
\end{figure*}

Our small sample of 7 bright mid-late M dwarf stars enabled one to two sets of observations to be made on each night. An example of the deconvolved profiles, plotted for GJ 1002, is shown in Fig. \ref{fig:profiles}. To monitor stability and reproducibility of results, we observed each target twice in a single pointing, with an intermediate ThAr frame, as discussed above. After extraction, we obtained S/N ratios of $\sim$25\,-\,160, which after deconvolution yielded single lines with S/N ratios of 875\,-\,7400. The vital statistics of each star, including the total number of observations are listed in Table \ref{tab:targets}.
The radial velocities derived from our targets are { listed in Table \ref{tab:rvs} and plotted in Fig. \ref{fig:rvs}}. Statistics, pertaining to precision of the measurements are given in Table \ref{tab:precision}. We are confident that the error estimates from {\sc hcross} are a good representation of the cross-correlation uncertainties. This is illustrated by columns 4 and 5 which give the {\sc hcross} error and the mean difference between pairs of observations for each object. In other words, over the short timescales on which pairs of observations are made (with no slewing of the telescope), the scatter in the points agrees well with the cross-correlation error estimate. While this also holds for GJ 3128 when the last observation pair (with a difference of 42 \ms) is excluded, the much higher \vsini\ and low S/N ratio of LP 944-20 may be the cause of the breakdown of this one-to-one relationship. There are also only two {\em pairs} of observations of LP 944-20 since the first was a single observation. 

In Table \ref{tab:precision}, the final column lists the discrepancy between the {\sc hcross} error and the scatter. The value is simply that which added in quadrature to the {\sc hcross} error leads to the observed scatter in the observations. As such it represents any further unaccounted for errors in the measurement, which may be astrophysical, instrumental, a real planetary signal, or a combination of these factors. In the next section we discuss such possible causes further.

The flattest RV curve is exhibited by GJ 1061, with an overall r.m.s. scatter of 15.7 \ms, comparable with the cross-correlation uncertainty of 9.37 \ms\ derived via {\sc hcross}. While the GJ 1286 RV curve is relatively flat, the overall scatter is more than twice the estimated error. In the case of GJ 1002, this discrepancy is closer to a factor of 3. Although LHS 132 appears relatively flat, we again emphasise that there are only two pairs of observations of this object, both taken at almost the same time on each night. No discernible variation is found in LP944-20 which shows the largest scatter and errors, owing to its late spectral type and high \vsini. To the contrary, GJ 3128 shows the same rising trend on each night of observations with a scatter that is $\sim$10 times the error estimate. Perhaps the most interesting RV curve is exhibited by SO J025300+165258, where a rising trend is seen on night one followed by a falling trend on the second night. We discuss GJ 3128 and SO J025300+165258 in more detail in section (\S \ref{section:candidates}).

\subsection{Photon noise sensitivities}
\protect\label{section:sensitivities}
To validate the technique we have carried out Monte-Carlo simulations to determine the radial velocity precision we expect to achieve as a function of S/N ratio. Synthetic spectra at R = 35,000 were generated from the semi-empirical line lists that we derived in \S \ref{section:linelists}. The spectra contained both telluric lines and spectral lines. An observed continuum (of GJ1002) was used to mimic the counts across the whole spectrum, thereby accounting for the spectral type and instrumental response (e.g. quantum efficiency and blaze function) of {\sc mike}at R $\sim$35,000. Noise was then added to the spectrum, and subjected to out pipeline procedure for deriving radial velocities. This procedure was carried out 100 times at each S/N level, resulting in 100 radial velocities from which the scatter, indicating the measurement precision was derived. The results are shown for mid-M dwarfs in Fig. \ref{fig:sn_simulation} for non-rotating stars. For the S/N range of 25\,-\,160 seen in our sample of stars, the photon limited precision is 1.7\,-\,11 \ms\ respectively. The right panel of Fig. \ref{fig:sn_simulation} demonstrates that with 10 \ms\ precision (for example by tellurics), there is little benefit in obtaining spectra with S/N ratios $>$ 100. 

{The lower curves plotted in Fig. \ref{fig:sn_simulation} (line joined by open circles) show the precision attainable from tellurics alone, indicating that they possess the larger contribution for low \vsini\ (i.e \vsini\ $\leq$ 5 \kms). For zero rotation \hbox{(\vsini\ = 0 \kms)}, the telluric cross-correlation uncertainties in the simulation are $\sim$ 1.5 times the stellar uncertainties. For example, with \hbox{S/N = 120} and \hbox{\vsini\ = 0 \kms}, the telluric and stellar contributions are 1.5 \ms\ and 2.2 \ms\ respectively. The combined precision of \hbox{2.6 \ms}\ is similar to that achieved using an I$_2$ cell and {\sc vlt-ut2 + uves} for the Barnard's star (M4V) \citep{kurster03barnard} and Proxima Centauri (M5V) \citep{endl08proxcen}. We note however that both objects possess earlier spectral types and additionally, were observed using an image slicer at a much higher resolution of R $\sim$ 100,000. A direct comparison of precision and efficiency (see also \S \ref{section:fluxes}) requires further simulations at other wavelengths to be made, although increased resolution would clearly improve our simulated precision further. We reiterate that telluric stability will nevertheless ultimately limit precision to a few \ms.}

\section{Discussion}
\protect\label{section:discussion}

\begin{figure}
\begin{center}
\includegraphics[width=60mm,angle=270]{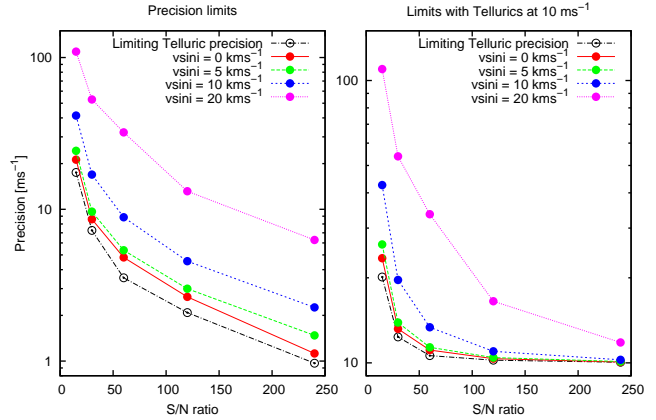} \\
\end{center}   
\caption{Left: Simulation showing the Poisson noise limited precision of our radial velocity method applied to {\sc MIKE} for 0.62\,-\,0.90 \micron\ with R = 35,000. { The simulation includes both the stellar and telluric line contributions.} The curves are plotted for stellar rotation of \vsini\ = 0, 5, 10 \& \hbox{20 \kms}. Right: The same curves showing the simulation with a \hbox{10 \ms}\ telluric uncertainty added in quadrature. { In both panels the lowest curve (open circles) indicates the limiting precision which is set by the telluric lines.}}
\protect\label{fig:sn_simulation}
\end{figure}

While precision of a few 10s of \ms\ has been demonstrated by our results presented in Fig. \ref{fig:rvs}, it is clear that the observations do not reach the Poisson noise limit. Our simulations (Fig. \ref{fig:sn_simulation}) indicate a photon noise limit of order 2.4 \ms\ for our brightest target, GJ 1061 (i.e. at S/N = 163). For GJ 1002 (S/N = 113), the photon noise limit is 3.3 \ms\ and similarly for GJ 1286 and GJ 3128 (S/N = 84), the limit is 4.4 \ms. For a modest S/N = 25, as for LHS 132, we expect a limiting precision of 14.5 \ms. With the exception of this latter case, for which we anyhow only have two observations, the photon-noise limit is clearly not sufficient to explain the discrepancy between the r.m.s. scatter in the points and the error bars listed in Table \ref{tab:precision}. LP944-20 is a rapid rotator with \vsini\ $\sim$ 35 \kms\ and our estimated photon noise limit (not shown in Table \ref{tab:precision}) at S/N = 44 is 99 \ms\ and therefore in reasonable agreement with both the error and scatter, which is smaller than the difference between pairs of observations. More data is needed on this kind of object which has a high \vsini (Jenkins et al., In prep.), even among M dwarfs \citep{jenkins09mdwarfs}.

\begin{table*}
 \begin{tabular}{lcccccc}
  Object		& S/N &  Error		& Mean $\Delta$OP	& r.m.s. scatter& Discrepancy &	Photon limited\\
			&     &  [ms$^{-1}$]	& [ms$^{-1}$]  		& [ms$^{-1}$]	& [ms$^{-1}$] & precision [ms$^{-1}$] \\
  \hline
GJ 1002			& 113 &   12.8		& 12.7			& 32.4		& 29.7 	&	3.3 \\
GJ 1061			& 163 &   9.37		& 11.8			& 15.7  	& 12.6 	&	2.5 \\
GJ 1286			&  84 &   8.63		& 8.88			& 22.1   	& 20.3 	&	4.4 \\ 
GJ 3128			&  84 &   8.80		& 20.0(12.6)		& 87.7  	& 87.2 	&	4.4 \\
SO J025300+165258	&  90 &   9.91		& 7.0			& 63.3   	& 62.5 	&	4.2 \\
LHS 132			&  25 &   29.7		& 27.2			& 33.7  	& 16.0 	&	14.5 \\
LP944-20		&  44 &   69.1		& 208			& 121		&99.3	&	7.6 \\
  \hline
 \end{tabular}
\caption{Radial velocity errors for each target. The third column is the error derived from the cross-correlation analysis. Column 4 ($\Delta$OP) is the mean uncertainty of the pairs of radial velocity points. Column 5 is the r.m.s. scatter of all observations for the object indicated. The value in brackets for GJ 3128 excludes the final pair which result in the higher value of 20.0 when all pairs are considered (see also Fig. 4). Column 6 lists the discrepancy between the scatter and the errors (columns 3 and 4), indicating the magnitude of other radial velocity components (i.e. planetary signal or noise) and column 7 gives the photon limited precision with {\sc mike}.}
\protect\label{tab:precision}
\end{table*}

The question of telluric stability has been tackled in a number of publications. Measurements of Arcturus and Procyon were made with $\sim$50 \ms\ precision by \cite{griffin73}. A number of authors, including \cite{snellen04} have demonstrated that sub-10 \ms\ is achievable by observing H$_2$O lines at wavelengths longer than 0.6 \micron. \cite{gray06tellurics} however found \hbox{25 \ms}\ precision from telluric observations of $\tau$ Ceti. More recently \cite{figueira10stability} have investigated the use of telluric features over timescales of a few days to several years by measuring the stability of some of the stronger O$_2$ lines found in {\sc harps} observations. Their observations indicate that for bright targets such as Tau Ceti, precision of $\sim$ 5 \ms\ is achievable on timescales of 1\,-\,8 d when observing at airmasses restricted to $<1.5$ (altitude $>$42\degs). Inclusion of data at all air masses ($<$2.2, altitude $>$ 27\degs) enables precision of $\sim 10$ \ms\ to be achieved. Over longer timescales of 1\,-\,6 years, the same precision of $\sim 10$ \ms\ is found. In other words, atmospheric phenomena are found to induce radial velocity variations in telluric lines at the 1\,-\,10 \ms\ level, but can largely be corrected for. Moreover, and importantly, simple fitting of atmospheric effects, including asymmetries found in molecular lines that vary as a function of altitude, and wind speed, enable corrections down to $\sim$ 2 \ms\ to be made. Figueira et al. find that this is twice the photon noise limit of their observations. If we include a 10 \ms\ uncertainty into our error budget, our overall contribution from tellurics and photon noise ranges from 10.2 \ms\ (GJ 1002) to 17.6 \ms\ (LHS 132). This simple argument is almost sufficient to declare the radial velocity curve of GJ 1061 to be flat within our uncertainties since the estimated discrepancy listed in Table \ref{tab:precision} is 12.6 \ms. Again, the same argument may be made for LHS 132, but clearly more observations are needed. The remaining M5.5V stars require modest (GJ 1286) to significant (GJ 1002) additional uncertainties to explain their discrepancies, while GJ 3128 and SO J025300+165258 both show more significant discrepancies. It is additionally worth noting that all observations of GJ 3128 were observed in the airmass range 1.38\,-1.43 (44\degs\,-\,46\degs), while SO J025300+165258 was observed with an airmass range of 1.49\,-1.71 (36\degs\,-\,42\degs), so airmass effects are unlikely to lead to radial velocity variations of order $\sim$ 100 \ms\ (see \S \ref{section:candidates}).

To investigate further the expected errors that may arise from molecular asymmetries in telluric lines, we used {\sc lbltrm} to simulate telluric spectra at an airmass range of 1.0\,-\,2.2. At R = 35,000, we find only $\sim$ 1 \ms\ variation for observations made at Las Campanas. The result additionally gives a measure of any uncertainty from using a mismatched line list for deconvolution. All our results have used a master telluric line list for an altitude 60\degs. Hence we do not expect airmass corrections to be important until we are confident that we have reached precision of a few \ms. A more important consideration potentially arises from wind speed and direction. Since most of the lines in our line list are water transitions, we expect the lines to form at lower altitudes. The soundings used (see \S \ref{section:linelists}) by {\sc lbltrm} give wind direction and speed for altitudes in the range $\sim$2\,-\,27 \kms. Since the water lines form largely at lower altitude, we note that the wind speed over the two nights of observations is $<$ 10 \ms\ on both nights. Additionally, the wind direction changes by only $\sim$20\degs, giving a maximum difference between the two nights of $\sim$ 0.6 \ms. With a wind vector of $360$\degs\ (Northerly), at altitudes $<$5000 m, the maximum wind speed seen by a star at altitude 40\degs\ in a 10 \ms\ wind is 10cos(40\degs) = 7.7 \ms\ during our two night observing run. Of course in reality, the differential is somewhat less since high declination stars only traverse a small range of azimuth angles. For a northerly wind at Southern latitudes, intermediate latitude stars traverse a greater range of azimuth angles, but will never see as much as the predicted full 7.7 \ms\ variation. We are therefore confident that in the absence of high precision measurements, telluric velocity corrections can not give the radial velocity variations seen in some of our stars. 

As {\sc mike} does not have an image derotator it is not able to keep the slit at the parallactic angle (it is set such that the slit is vertical at 30\degs\ from the Zenith (altitude = 60\degs). It is unclear whether this should have an effect on the measured radial velocities. One might reasonably expect that to first order, any slit effects would cancel owing to the reference lines and stellar lines following the same optical path in the instrument. There is no clear systematic correction that can be applied to all the objects that appears to correct, in an empirical manner, for the slit tilt.

\subsection{Planetary candidates?}
\protect\label{section:candidates}

\begin{figure}
\begin{center}
\includegraphics[width=79mm,angle=270]{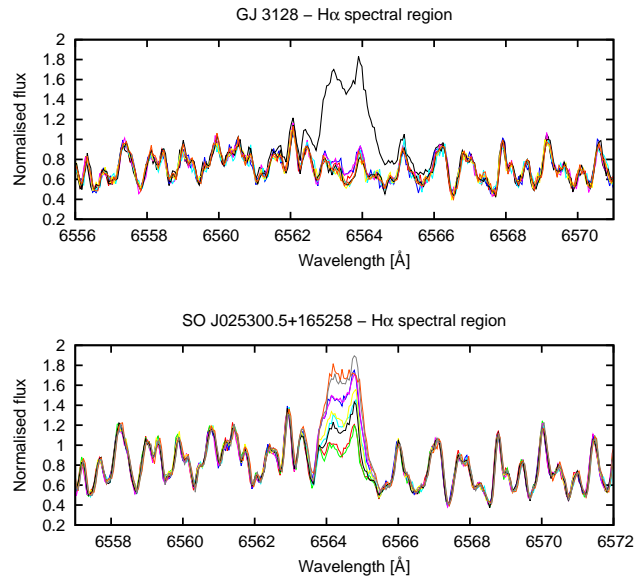} \\
\end{center}   
\caption{The normalised spectral region around \ha\ for all GJ 3128 and \hbox{SO J025300.5+165258} spectra. For GJ 3128, the second observation from the first night shows clear evidence of a flaring event which has pushed \ha\ into emission. SO J025300.5+165258 shows that \ha\ is variable from one spectrum to the next, over all 9 spectra. No other spectra show this degree of variability.}
\protect\label{fig:halpha}
\end{figure}

Although GJ 3128 appears to show significant radial velocity variations that are at 10$\sigma$ of the cross-correlation errors, the same ascending RVs of both observation pairs on both nights raises the possibility of an as yet unidentified one day aliasing effect. The radial velocity points can be fit with a simple sinusoidal curve with a period of $P$ = 0.511 d, an amplitude of $K_*$ = 209.4 \ms\ and a residual r.m.s. of 1.8 \ms. A circular \hbox{(eccentricity, e = 0)} planetary orbit of this period and amplitude could be induced by a \hbox{$m_p$~sin~$i$ = 3.29 M$_{\rm Nep}$} planet in a very close 0.0058 AU orbit (i.e. 8.31 R$_*$). We note also that under the assumption of a maximum \vsini\ = 5 \kms, the minimum period of the GJ 3128, with \hbox{R = 0.15 \rsun}\ \citep{reid05,kalteneggar09}, would be P $\sim$1.52 d, quite distinct from the RV signal. Importantly, Fig. \ref{fig:halpha} shows that GJ 3128 exhibits flaring activity. Our second observation shows clear evidence of such an event which has pushed \ha\ into emission. This interesting phenomenon confirms the findings of \cite{reiners09flare}, already discussed in \S 1, and demonstrates that we are not sensitive to flaring events at the level of our precision since the first pair of points do not show significant radial velocity differences. Moreover, since no further strong flaring is seen in the remaining spectra, we believe that such events should not be responsible for the RV variation seen in GJ 3128. We also emphasise that we exclude a region of $\pm$ 7 \AA\ surrounding \ha\ from our deconvolution.

Our most promising candidate for a stellar radial velocity signature is exhibited by SO J025300.5+165258, which shows a rising trend on the first night and a falling trend on the second night. The r.m.s. scatter is at a level of 6.4$\sigma$ of the cross-correlation errors. A sinusoidal fit to the data indicates a period of \hbox{$P$ = 2.06 d} with an amplitude of \hbox{$K_*$ = 182.7 \kms}\ (residual r.m.s. = 0.46 \ms). A circular planetary orbit would lead to a mass of \hbox{$m_p$~sin~$i$ = 4.88 M$_{\rm Nep}$} with \hbox{a = 0.014 AU}, falling inside the classical habitable zone. We emphasise that we make no claims for a planet and clearly further observations are required. In fact, perhaps Fig. \ref{fig:halpha} shows that in fact \hbox{SO J025300.5+165258} is the most chromospherically active star in our sample, but on the basis of our GJ 3128, the \cite{reiners09flare} results and removal of these regions from the RV analysis, it seems unlikely that these chromospheric variations are responsible for the RV variations of \hbox{SO J025300.5+165258}. However, while pairs or triplets of observations are consistent at the cross-correlation level, the \ha\ line also does not show large variations on these timescales, but {\em does} in fact vary from one pair/triplet to the next. Additionally, for completeness, we note that no other stars exhibit significant variations in the \ha\ line region, except for the final pair of observations of \hbox{GJ 1286}, which show a slight increase \ha\ emission.

\subsection{Line bisectors}
\protect\label{section:bisectors}

\begin{figure}
\begin{center}
\includegraphics[width=60mm,angle=270]{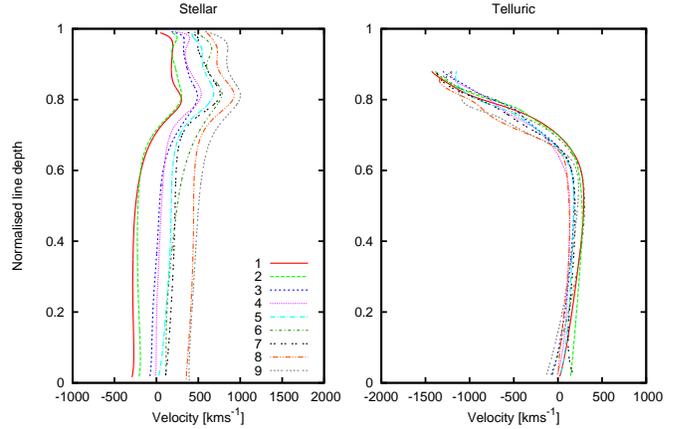} \\
\end{center}   
\caption{Line bisector plots for all 9 observations of SO J025300.5+165258 for the deconvolved stellar (left) lines and telluric lines (right). We are unable to determine a clear bisector span vs RV correlation, but note that we observe 143 \ms\ variation in the stellar bisector span.}
\protect\label{fig:bisplot}
\end{figure}

Although we believe that chromospherically active lines themselves do not bias our radial velocities, the active nature of the star clearly warrants further analysis of activity indicators. Line bisectors are a commonly employed tool that enable line asymmetries due to photospheric temperature inhomogeneities such as starspots and/or plage to be assessed \citep{saar97spots}. Following the same procedure as \cite{fiorenzano05bisectors}, we calculate line bisectors to assess starspot jitter. Fig. \ref{fig:bisplot} shows our results for \hbox{SO J025300.5+165258} for both the deconvolved stellar lines and telluric lines. The line bisectors are shown for all 9 profiles. The bisector span is calculated by determining the mean bisector for two regions of the line profiles, near the top and the bottom of the profile. Thus, any spot or spot group on the star that rotates into and out of view will cause variation in the span, with the subsequent line asymmetries yielding false RV signals (as defined, the bisector span, BIS, shows an anti-correlation with RV \citep{desort07rvs}). The telluric line profiles are quite asymmetric in the line wings, but are relatively consistent over all observations. The stellar lines show more variation in the wings (i.e. at normalised depths $>$0.6), but even points 5-7 (the triplet of observations on night 2 which show bisector morphology changes in short timescales), do not show large variation in their RVs. 

With so few observations, we find that it is in fact not possible to find a bisector-span vs RV correlation owing to the scatter in the bisector values. We combine bisector span values from both the telluric and stellar lines since this should also account for any changes in the telluric lines. With a simple linear relationship, it {\em is} possible to flatten the two pairs of points on night 1, but this has little effect on the points from night 2. It is clear that a larger number of data points are needed at more phases before any bisector span vs radial velocity relationship can provide useful information on the nature of radial velocity variations. However, with a total variation in the bisector span of $\sim$150 \ms\ for \hbox{SO J025300.5+165258} (92 \ms\ when the large span calculate for profile 6 is removed), we can't completely rule out that a trend would emerge with more observations. We also find similar bisector variations in our other stars, including our brightest target, GJ1061, which is nevertheless relatively flat within the estimated cross-correlation uncertainties. It is worth noting that the $\sim$1 per cent FWHM variations that we see in our profiles also do not yield a clear (anti-)correlation, as reported by \cite{boisse11}. As with \hbox{SO J025300.5+165258}, we are unable to determine a correlation for GJ 3128, but do find that the second observation (where we reported a flare, as shown if Fig. \ref{fig:halpha}), is consistent with the first observation, confirming that it had little effect on the photospheric lines. The above findings may well be a consequence of lower contrast starspots in low temperature atmospheres that contribute negligible stellar jitter, a hypothesis that may yield clearer answers with further data.

\subsection{Detection analysis and limits}
\protect\label{section:detectionlimits}

In the case of no planetary RV variations, we are sensitive (at 1$\sigma$ scatter) down to planet masses of 6 \mearth\ in the case of GJ 1061. Previous simulations \citep{barnes11jitter} in the near infrared Y-band have shown that only of order 20 epochs are required to detect 5 \mearth\ planets orbiting 0.1 \msun\ stars rotating with \vsini = 5 \kms\ and exhibiting solar-like spot activity. The simulations presented in  \cite{barnes11jitter} used standard periodogram analyses to search for planetary signals in fake radial velocity curves injected with astrophysical starspot jitter. However an analysis of several marginal detections and non-detections using Bayesian techniques that are efficient at searching for low amplitude signals in noisy data \citep{kotiranta10,tuomi11,tuomi12} indicate that the number of epochs required is conservative. Clustering of randomly located starspots in our 3D stellar models was found to lead to non-symmetric noise/jitter distributions using these routines. In some instances, using the correct noise-model resulted in the recovery of orbital parameters, where no planet had even been detected at the 1 per cent false alarm probability level of our more straightforward periodogram analysis. This observation may well prove vital for stellar systems where the rotation period of the star may potentially be close to that of the planet, a distinct possibility for HZ planets orbiting mid-late M stars.

\section{Summary \& Prospects}
\protect\label{section:conclusion}

By making use of the red optical spectral regions and an efficient CCD detector we have demonstrated that sub-20 \ms\ precision radial velocity measurements of M dwarfs are possible when utilising a telluric reference fiducial and a wavelength extent \hbox{($\sim$3000 \AA)} equal to more traditional optical surveys. Two stars show significant radial velocity variations that are consistent with the RV amplitudes that could be induced by close orbiting planets of several Neptune masses. At this stage, with so few observation epochs, we do not argue that these are planetary in origin, although we note that Neptune mass planets are expected \citep{howard10,wittenmyer11,bonfils11mdwarfs,borucki11kepler} with a frequency that is consistent with detection rates of 1/7\,-\,2/7. While we find that the chromospheric activity, seen as variation in \ha\ emission, is unlikely to be the cause of the variations, based on local consistency of observation pairs (e.g. where one observation of a pair contains clear evidence for a strong flare) and other findings \citep{reiners09flare}, further observations are required to enable a fuller investigation aimed at the role of photospheric line shape variability.

Using simulations based on our {\sc mike} observations, we have shown that sub-10 \ms\ precision can potentially be achieved for spectra with S/N $\sim$50. This improves to precisions of order 3 \ms\ for S/N $\sim$ 100 and 2 \ms\ for S/N $\sim$ 200. Since rotation is known to increase on average for mid-late M dwarfs, more modest precisions of 5\,-\,12 \ms\ are more likely once rotation speeds reach typical 10\,-\,20 \ms\ for M6V and M9V stars respectively \citep{jenkins09mdwarfs}.

Our observations suggest that we have achieved reference fiducial limited observations of at least one target, GJ 1061, where the discrepancy between our cross-correlation errors, which are accurate on short timescales, and the overall scatter, can be explained by atmospheric variations from tellurics. Since these should in fact be sub-10 \ms\ during our observations, there are likely additional uncertainties which most probably arise from {\sc mike}. We have demonstrated that the instrument does not show a slow velocity drift with time, but rather that apparent drifts of a few hundred \ms\ can be observed on relatively short timescales of a few observations. We should add that {\sc mike} was not built with precision radial velocities in mind and is not pressure or temperature stabilised, although the Magellan Planet Search program which operates in the optical and utilises an iodine cell is achieving 5 \ms\ precision \citep{minniti09lowmass,arriagada10magellan}. It is not clear whether a mechanical effect, possibly introduced by the calibration mirror while taking arcs, or whether telescope slewing and possible disturbance of the Nasmyth mounted instrument are responsible for these shifts. Additionally, we are unable to determine whether the shifts are pseudo-random and occur only during movement of the telescope and/or instrument components, or whether the influence of slow temperature and pressure changes throughout the night are important. If gravity settling of the instrument is significant (esp. after a CCD dewar refill), we would expect possible drifts of up to several 100 \ms\ during a single observation. Fig. \ref{fig:stabilityplot} appears to indicate that the shifts are not entirely random as they often continue in one direction over several observations. As discussed in the previous section for the variable slit angle, owing to the simultaneous measurement of reference fiducial and stellar spectrum, which traverse the same optical light path, such effects may well be expected to cancel to first order. Nevertheless, they may go some way to explaining the discrepancy between residuals and our error bar estimates.

In light of such difficulties, {\sc mike}, operating at a modest resolution of $\sim$35,000 (especially when we consider that the temperature and pressure stabilised {\sc harps} operates at R = 110,000) has demonstrated that precision radial velocities are well within its grasp. This work would nevertheless clearly benefit from a more stable instrument operating at higher resolution. We expect our precision to scale approximately linearly with resolution, so the above estimates of photon limited noise, would potentially enable 1 \ms\ precision to be achieved on the brightest slowly rotating M dwarfs at a resolution of R $\sim$50,000. The additional S/N loss at this resolution could easily be offset by telescopes with a larger aperture than the 6.5m Magellan Telescopes could offer. While we have not been able to determine the effects of a slit that does not maintain parallactic angle on the sky, we expect that doing so could only add to the precision (even if reddening effects at low airmass are reduced in the re-optical). Instruments such as UVES have already demonstrated precision of a 2\,-\,2.5 \ms\ over 7 years using an iodine cell \citep{zechmeister09uves} and additionally offer greater red sensitivity. This opens up the potential of using the 0.9\,-\,1.0 \micron\ wavelength region. The richness of atmospheric lines over most of this range, combined with the relatively free z-band region just short of \hbox{1.0 \micron}, and where M dwarf fluxes are even greater, would clearly be of great benefit to this kind of study.

We see no reason why the 2 \ms\ precision reported by \cite{figueira10stability} can not be achieved on existing 8 m class telescopes with existing instruments. By spectral type M6V, habitable zone rocky planets of 1 \mearth\ would be expected to induce radial velocity variations of similar magnitude to this precision while the most massive rocky planets with 10 \mearth\ would be capable of detecting signals at levels of order 10$\sigma$ in $\leq$ 20 epochs of observations over short timescales of days.

\section*{Acknowledgments}
{We thank the referee for careful reading of the manuscript and for suggestions which have improved the final version.} JB gratefully acknowledges funding through a University of Hertfordshire Research Fellowship. JSJ acknowledges funding by Fondecyt through grant 3110004 and partial support from Centro de Astrof\'{i}sica FONDAP 15010003, the GEMINI-CONICYT FUND and from the Comit\'{e} Mixto ESO-GOBIERNO DE CHILE. {DM and PA gratefully acknowledge support by the FONDAP Center for Astrophysics 15010003, the BASAL CATA Center for Astrophysics and Associated Technologies PFB-06, and the MILENIO Milky Way Millennium Nucleus from the Ministry of Economy’s ICM grant P07-021-F. AJ acknowledges support from Fondecyt project 1095213, Ministry of Economy ICM Nuclei P07-021-F and P10-022-F, and Anillo ACT-086.} HRAJ, DJP and MT are supported by RoPACS, a Marie Curie Initial Training Network funded by the European Commission’s Seventh Framework Programme. JB, JSJ, DJP and SVJ have also received travel support from RoPACS during this research. This paper includes data gathered with the 6.5 meter Magellan Telescopes located at Las Campanas Observatory, Chile.


\protect\label{lastpage}
\end{document}